\documentclass[prl,aps,twocolumn,showpacs]{revtex4}
\usepackage{epsfig}
\usepackage{bm}

\begin{document}

\title{Subharmonic resonance of global climate to solar forcing}

\author{\small  A. Bershadskii}
\affiliation{\small {ICAR, P.O.B. 31155, Jerusalem 91000, Israel}}

\begin{abstract}
It is shown that, the wavelet regression detrended fluctuations 
of the monthly global temperature data (land and ocean combined) 
for the period 1880-2009yy, are completely dominated by one-third subharmonic 
resonance to annual forcing (both natural and anthropogenically
induced). Role of the oceanic Rossby waves and the 
resonance contribution to the $El~Ni\tilde{n}o$ phenomenon have been discussed 
in detail.
 
\end{abstract}


\maketitle

\section{Introduction}

The monthly global temperature data are known to be strongly fluctuating. 
Actually, the fluctuations are of the same order as the trend itself (see figures 1 and 2). 
While the nature of the trend is widely discussed (in relation to the global warming) the 
nature of these strong fluctuations is still quite obscure. The problem has also a technical 
aspect: detrending is a difficult task for such strong fluctuations. 
In order to solve this 
problem a wavelet regression detrending method was used in present investigation. Then a 
spectral analysis of the detrended data reveals rather surprising nature of the strong 
global temperature fluctuations. Namely, the detrended fluctuations of the global temperature 
for the last century (the instrumental monthly data are available at Ref. \cite{12data}, see also 
Ref. \cite{s}) are completely dominated by so-called one-third subharmonic resonance to annual 
forcing.  
     
\begin{figure} \vspace{-0.5cm}\centering
\epsfig{width=.45\textwidth,file=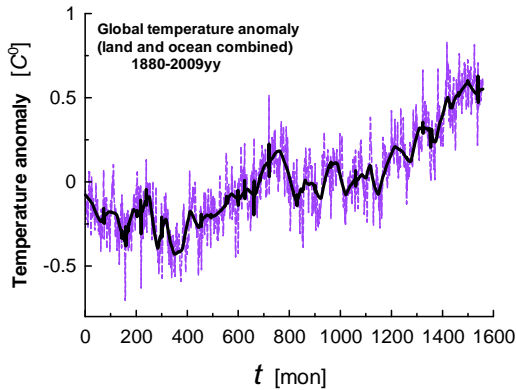} \vspace{-4.5cm}
\caption{The monthly global temperature data (dashed line) for the period 1880-2009. The solid curve 
(trend) corresponds to a wavelet (symmlet) regression of the data. }
\end{figure}
\begin{figure} \vspace{-0.5cm}\centering
\epsfig{width=.45\textwidth,file=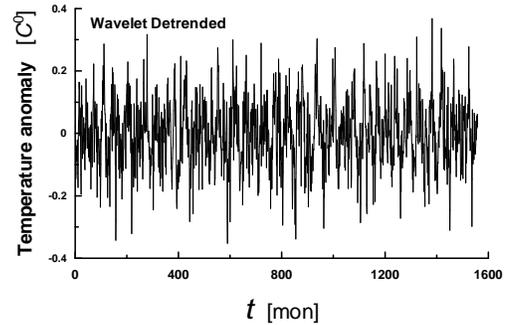} \vspace{-5.5cm}
\caption{The wavelet regression detrended fluctuations from the data shown in Fig. 1.}
\end{figure}
\begin{figure} \vspace{-0.5cm}\centering
\epsfig{width=.45\textwidth,file=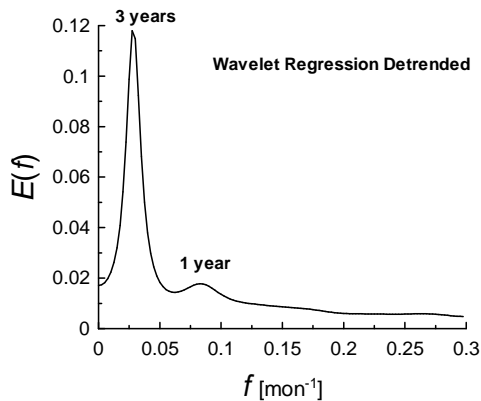} \vspace{-4.5cm}
\caption{Spectrum of the wavelet regression detrended fluctuations of 
the monthly global temperature anomaly (land and ocean combined).}
\end{figure}
\begin{figure} \vspace{-0.5cm}\centering
\epsfig{width=.45\textwidth,file=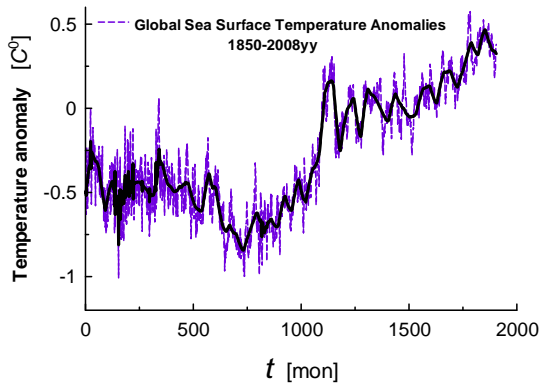} \vspace{-4.5cm}
\caption{The monthly global Sea Surface Temperature Anomalies (dashed line) for the period 1850-2008yy 
(the data taken from Ref.\cite{sea-data}). The solid curve (trend) corresponds to a wavelet 
(symmlet) regression of the data. }
\end{figure}

\begin{figure} \vspace{-0.5cm}\centering
\epsfig{width=.45\textwidth,file=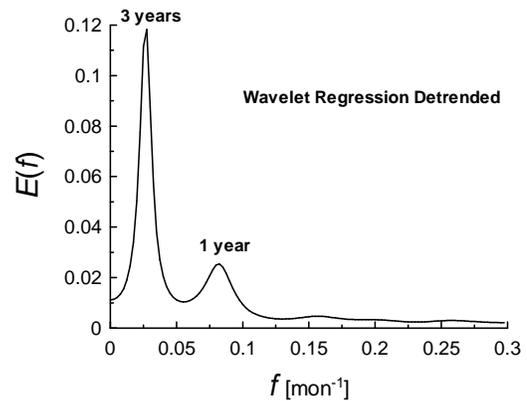} \vspace{-4.5cm}
\caption{The same as in Fig. 3 but for the wavelet regression detrended fluctuations of the global Sea Surface Temperature Anomalies.}
\end{figure}

\section{Subharmonic resonance to annual forcing}

There are many well known reasons for asymmetry in response of the North and South 
Hemispheres to solar forcing: dominance of water in the Southern Hemisphere against 
dominance of land in the Northern one, topographical imbalance of land (continents) and 
oceans in the Northern Hemisphere due to continental configuration, seasonality and vegetation 
changes are much more pronounce on land than on ocean surface, and anthropogenically induced asymmetry of the last century. This asymmetry results 
in {\it annual} asymmetry of global heat budget and, in particular, 
in annual fluctuations of the global temperature. Nonlinear responses are expected as a 
result of this asymmetry. 

   Figure 1 shows (as dashed line) the instrumental monthly global temperature data 
(land and ocean combined) for the period 1880-2009, as presented at the NOAA site \cite{12data}. 
The solid curve (trend) in the figure corresponds to a wavelet (symmlet) regression of the data 
(cf Refs. \cite{sw},\cite{o}). Figure 2 shows corresponding detrended fluctuations, which produce 
a statistically stationary set of data. Most of the regression methods are linear in responses. 
At the nonlinear nonparametric wavelet regression one chooses a relatively small number of wavelet 
coefficients to represent the underlying regression function. A threshold method is used to keep or 
kill the wavelet coefficients. In this case, in particular, the Universal (VisuShrink) thresholding 
rule with a soft thresholding function was used. At the wavelet 
regression the demands to smoothness of the function being estimated are relaxed considerably in comparison 
to the traditional methods. 
Figure 3 shows a spectrum of the wavelet regression detrended data 
calculated using the maximum entropy method (because it provides an optimal spectral
resolution even for small data sets). One can see in this figure a small peak corresponding 
to a one-year period and a huge well defined peak corresponding to a three-years period. 

  In order to understand underlying physics of the very characteristic picture shown in the Fig. 3 
let us imagine a forced excitable system with a large amount of loosely coupled degrees of freedom 
schematically represented by Duffing oscillators (which has become a classic model for analysis of 
nonlinear phenomena and can exhibit both deterministic and chaotic behavior \cite{ot}-\cite{b} 
depending on the parameters range) with a wide range of the 
natural frequencies $\omega_0$ (it is well known \cite{tw} that oscillations with a wide range of 
frequencies are supported by ocean and atmosphere, cf also Ref. \cite{bkb}):

$$
\ddot{x} + \omega_0^2 x +\gamma \dot{x} +\beta x^3 = F \sin\omega t    \eqno{(1)}
$$
where $\dot{x}$ denotes the temporal derivative of $x$, $\beta$ is the strength of nonlinearity, and 
$F$ and $\omega$ are characteristic of a driving force. It is known (see for instance Ref. \cite{nm}) 
that when $\omega \approx 3\omega_0$ and $\beta \ll 1$ the equation (1) has a resonant solution 
$$
x(t) \approx a \cos\left(\frac{\omega}{3}t + \varphi \right) + \frac{F}{(\omega^2-\omega_0^2)} 
\cos \omega t   \eqno{(2)}
$$
where the amplitude $a$ and the phase $\varphi$ are certain constants. 
This is so-called one-third subharmonic resonance with the driving frequency 
$\omega$ corresponding to the {\it annual} NS-asymmetry of the solar forcing 
(the huge peak in Fig. 3 corresponds to the first term in the right-hand side of the Eq. (2)). 
For the considered system of the oscillators an effect of synchronization can take place
and, as a consequence of this synchronization, the characteristic peaks in the spectra 
of partial oscillations coincide \cite{nl}. 
It can be useful to note, for the global climate modeling, that the odd-term subharmonic resonance 
is a consequence of the reflection symmetry of the natural nonlinear oscillators 
(invariance to the transformation $x \rightarrow -x$, cf. also Ref. \cite{mj}).

\begin{figure} \vspace{-1cm}\centering
\epsfig{width=.45\textwidth,file=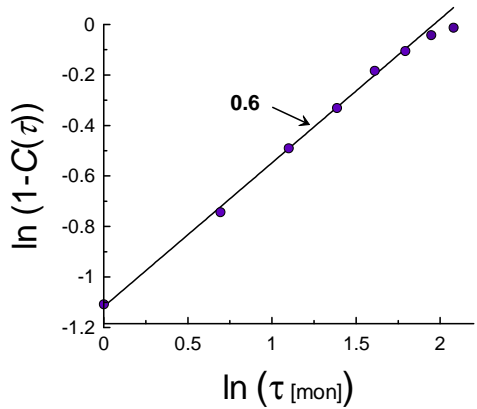} \vspace{-4cm}
\caption{Defect of autocorrelation function versus $\tau$ (in ln-ln scales) for the wavelet regression detrended fluctuations of the global Sea Surface Temperature Anomalies. The straight line is drawn in order to indicate 
scaling Eq. (3). }
\end{figure}
\begin{figure} \vspace{-0.5cm}\centering
\epsfig{width=.45\textwidth,file=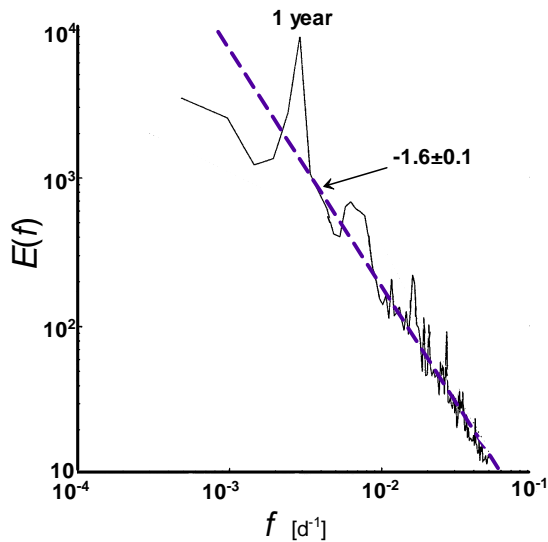} \vspace{-2cm}
\caption{Spectrum of sea surface height fluctuations \cite{zw} (TOPEX/
Poseidon and ERS-1/2 altimeter measurements) in the logarithmic scales. 
The profound peak corresponds to the annual cycle. 
The dashed straight line is drawn in order to indicate correspondence to the scaling Eq. (3).}
\end{figure}

\section{Role of Rossby waves}

The fluctuations of oceanic temperature cause certain variations of the sea surface height. 
These variations are intermixed with the sea surface height variations caused by the oceanic planetary Rossby waves. 
The oceanic planetary Rossby waves play an important role in the response of the global 
ocean to the forcing (see, for instance, Refs. \cite{pl},\cite{zw}) and they are of fundamental importance 
to ocean circulation on a wide range of time scales (it was also suggested that the Rossby 
waves play a crucial role in the initiation and termination of the $El~Ni\tilde{n}o$ phenomenon, see also below). 
Therefore, they present a favorable physical background for the global subharmonic resonance. 
For that reason it is interesting to look also separately at global sea surface temperature anomalies. 
These data for time range 1850-2008yy are shown in Fig. 4 (the monthly data were taken from 
Ref. \cite{sea-data}, see also Ref. \cite{s2}). The solid curve (trend) in the figure 
corresponds to the wavelet (symmlet) regression of the data.  
Figure 5 shows a spectrum of the wavelet regression detrended data 
calculated using the maximum entropy method. The spectrum seems to be 
very similar to the spectrum presented in Fig. 3. 

Since the high frequency part of the spectrum is corrupted by strong fluctuations (the Nyquist frequency 
equals 0.5 [$mon^{-1}$]), it is interesting to look at corresponding autocorrelation function $C(\tau)$ in order to understand what happens on the monthly scales. It should be noted that scaling of defect of the autocorrelation 
function can be related to scaling of corresponding spectrum:
$$
1-C(\tau) \sim \tau^{\alpha}~~~~~~\Leftrightarrow~~~~~~~~ E(f) \sim f^{-(1+\alpha)}   \eqno{(3)}
$$
Figure 6 shows the defect of the autocorrelation function in ln-ln scales in order to estimate the scaling 
exponent $\alpha \simeq 0.6 \pm 0.04$ (the straight line in this figure indicates the scaling Eq. (3)). 
The existence of oceanic Rossby waves was confirmed rather recently by NASA/CNES TOPEX/Poseidon satellite 
altimetry measurements. Corresponding to these measurements spectrum of the sea surface height 
fluctuations, calculated in Ref. \cite{zw} with {\it daily} resolution (see also \cite{zfw}), 
is shown in Figure 7. The dashed straight line in this figure is drawn in order to indicate 
correspondence to the scaling Eq. (3): $1+\alpha \simeq 1.6$.  

\begin{figure} \vspace{-0.5cm}\centering
\epsfig{width=.45\textwidth,file=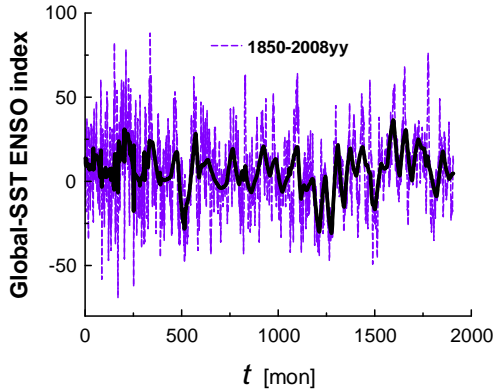} \vspace{-4.5cm}
\caption{The monthly Global-SST ENSO index (dashed line) for the period 1850-2008yy (the data taken from Ref. \cite{sst-enso}, the index is in hundredths of a degree Celsius) . The solid curve (trend) corresponds to a wavelet (symmlet) regression of the data. }
\end{figure}

\begin{figure} \vspace{-0.5cm}\centering
\epsfig{width=.45\textwidth,file=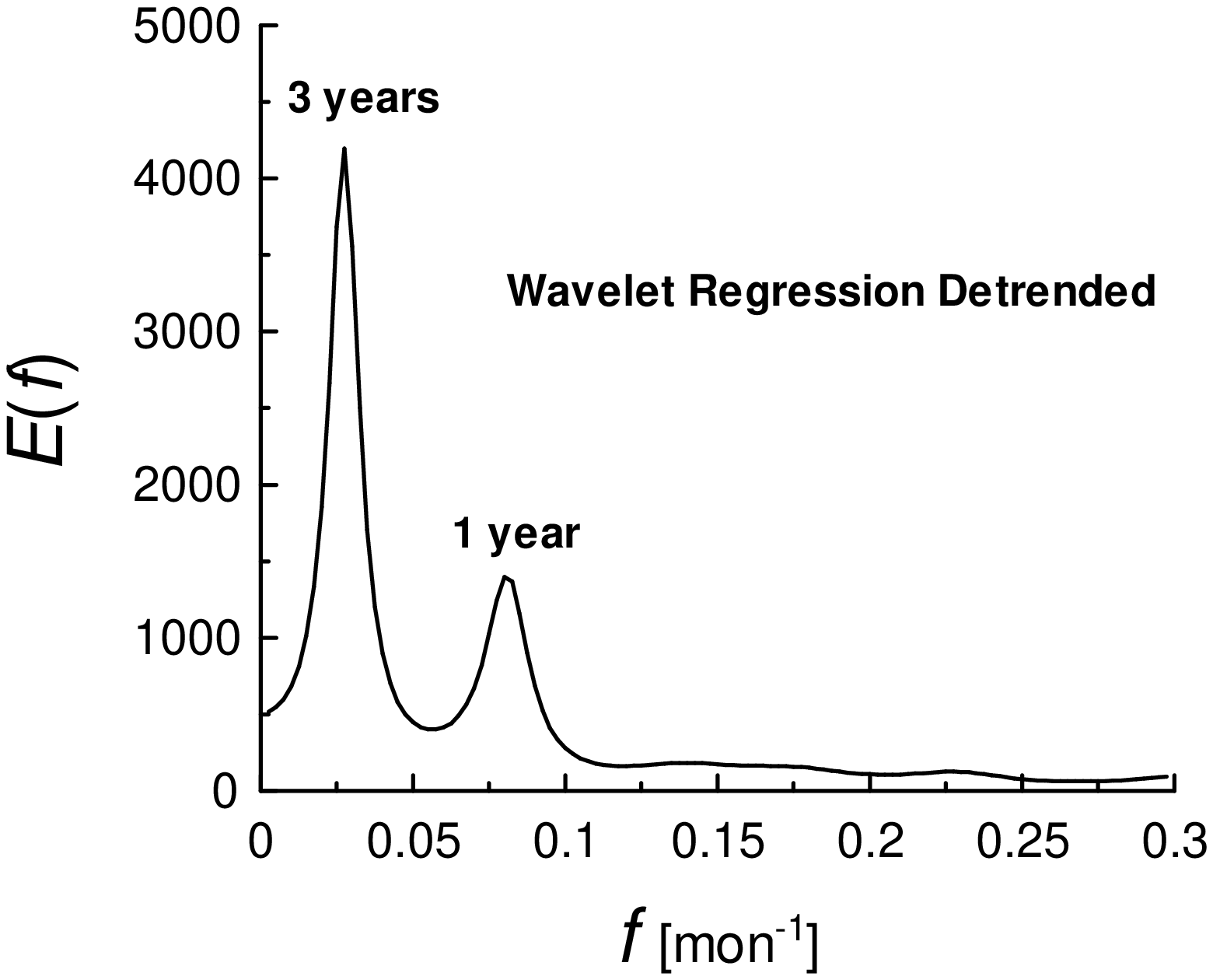} \vspace{-4.5cm}
\caption{The same as in Fig. 3 but for the wavelet regression detrended fluctuations 
of the Global-SST ENSO index.}
\end{figure}

\begin{figure} \vspace{-0.5cm}\centering
\epsfig{width=.45\textwidth,file=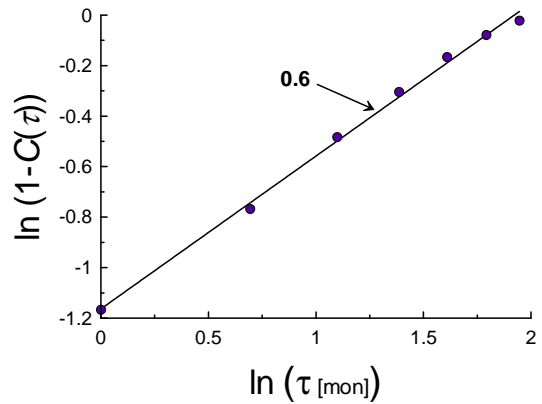} \vspace{-4.5cm}
\caption{The same as in Fig. 6 but for the wavelet regression detrended fluctuations 
of the Global-SST ENSO index. }
\end{figure}

\section{El~Nino phenomenon}

The Rossby waves (together with Kelvin waves) and a strong atmosphere-ocean feedback provide physical 
background for the $El~Ni\tilde{n}o$ phenomenon (see, for instance, Ref. \cite{tzi} and references therein). 
Figure 9 shows spectrum for the wavelet detrended fluctuations 
of the so-called Global-SST ENSO index (Fig. 8), which captures the low-frequency part 
of the $El~Ni\tilde{n}o$ phenomenon (the monthly data are available at Ref. \cite{sst-enso}). The annual 
forcing can come from the oceanic Rossby waves (cf Fig. 7). 
To support this relationship we show in figure 10 defect of autocorrelation function calculated using the 
wavelet detrended fluctuations from Fig. 8. 
The ln-ln scales have been used in Fig. 10 in order to estimate the scaling 
exponent $\alpha \simeq 0.6 \pm 0.03$ (the straight line in this figure indicates the scaling Eq. (3), 
cf Figs. 6 and 7). Using these observations one can suggest that the $El~Ni\tilde{n}o$ phenomenon 
has the one-third subharmonic resonance as a background. \\

The data were provided by National Climatic Data Center at NOAA and by Joint Institute for the Study
of the Atmosphere and Ocean. I also acknowledge that a software provided 
by K. Yoshioka was used at the computations.


\begin{thebibliography}{99}
\bibitem{12data} The data are available at http://lwf.ncdc.noaa.gov/

oa/climate/research/anomalies/index.html
\bibitem{s} T.M. Smith, et al., Improvements to NOAA's Historical Merged Land-Ocean Surface Temperature Analysis (1880-2006), J. Climate, {\bf 21}, 2283 (2008).
\bibitem{sw} N. Scafetta, and B. J. West, Phenomenological solar contribution to
the 1900-2000 global surface warming, Geophys. Res. Lett., {\bf 33}, L05708 (2006).
\bibitem{o} T. Ogden, Essential Wavelets for Statistical Applications and Data Analysis 
(Birkhauser, Basel, 1997).
\bibitem{ot} E. Ott, Chaos in Dynamical Systems (Cambridge University Press, 2002).
\bibitem{ph} D. Permann and I. Hamilton, Wavelet analysis of time series for the Duffing oscillator: 
The detection of order within chaos, Phys. Rev. Lett., {\bf 69}, 2607 (1992).
\bibitem{bh} V. Brunsden and P. Holmes, Power spectra of strange attractors near homoclinic orbits, 
Phys. Rev. Lett., {\bf 58}, 1699 (1987). 
\bibitem{b} A. Bershadskii, Chaotic climate response to long-term solar forcing variability, EPL
(Europhys. Lett.), {\bf 88}, 60004 (2009).
\bibitem{tw} S.M. Tobias, and N.O. Weiss, Resonant Interactions between Solar Activity and Climate, 
J. Climate, {\bf 13}, 3745 (2000). 
\bibitem{bkb} J. Brindley, T. Kapitaniak, and A. Barcilon, 
Chaos and noisy periodicity in forced ocean-atmosphere models, Phys. Lett. A, {\bf 167}, 179-184 (1992). 
\bibitem{nm} A.H. Nayfeh and D.T. Mook, "Nonlinear Oscillations" (John Wiley \& Sons, 
a Wiley-Interscience Publication, 1979).
\bibitem{nl} Yu.I. Neimark and P.S. Landa, Stochastic and Chaotic Oscillations,
(Dordrecht, Kluwer, 1992).
\bibitem{mj} S. Minobe, and F-f Jin., Generation of interannual and interdecadal climate oscillations
through nonlinear subharmonic resonance in delayed oscillators, Geophys. Res. Lett., {\bf 31}, 
L16206, (2004).
\bibitem{pl} P.S. Polito, and W.T. Liu, Global characterization of Rossby waves at several 
spectral bands, J. Geophys. Res. - Oceans, {\bf 108}, 3018 (2003).
\bibitem{zw} X. Zang and C. Wunsch, Spectral description
of low frequency oceanic variability, J. Phys. Oceanogr., {\bf 31} 3073 (2001).
\bibitem{zfw} X. Zang, L.L. Fu, and C. Wunsch, Observed reflectivity of
the western boundary of the equatorial Pacific Ocean. J. Geophys.
Res., {\bf 107}, 3150 (2002).
\bibitem{sea-data} The data are available at http://jisao.washington.edu

/data/global\_sstanomts/
\bibitem{s2} T.M. Smith, et al., Reconstruction of historical sea surface temperatures using 
empirical orthogonal functions, J. Climate, {\bf 9}, 1403 (1996).
\bibitem{tzi} E. Tziperman, H. Scher, S.E. Zebiak and M. A. Cane, Controlling Spatiotemporal 
Chaos in a Realistic El Nino Prediction Model, Phys. Rev. Lett., {\bf 79}, 1034-37 (1997).
\bibitem{sst-enso} The data are available at http://jisao.washington.edu/

data/globalsstenso/

\end{thebibliography}
\end{document}